\documentclass[aip,sd,amsmath,amssymb,reprint]{revtex4-1}

\usepackage{color}
\usepackage{graphicx}
\usepackage{dcolumn}
\usepackage{bm}
\usepackage{graphicx}
\usepackage{amsmath}
\usepackage{mathptm} 
\usepackage{hyperref}
\preprint{APS/123-QED}
\usepackage[normalem]{ulem}
\usepackage{xcolor}

\begin{document}

\preprint{APS/123-QED}

\title{Dangerous  Aging Transition in a Network of Coupled  Oscillators }
\author{Biswambhar Rakshit}%
 \email{biswambhar.rakshit@gmail.com}
\affiliation{Department of Mathematics, Amrita School of Engineering, Coimbatore, Amrita Vishwa Vidyapeetham, $641 112$, India}%

\author{Niveditha Rajendrakumar}
\email{niveditha.rajendrakumar@gmail.com}
 \affiliation{Aerodynamics and Wind Energy Department,  Faculty of Aerospace Engineering, Delft University of Technology} 
  \author{Bipin Balaram}%
 \email{b\_bipin@cb.amrita.edu}
\affiliation{Department of Mechanical Engineering, Amrita School of Engineering, Coimbatore, Amrita Vishwa Vidyapeetham, $641 112$, India}%

\date{\today}

\begin{abstract}
In this article, we investigate the dynamical robustness in a network of relaxation oscillators. In particular, we consider a network of diffusively coupled Van der Pol oscillators to explore the aging transition phenomena. Our investigation reveals that the mechanism of aging transition in a network of Van der Pol oscillator is quite different from that of typical sinusoidal oscillators such as Stuart-Landau oscillators. Unlike sinusoidal oscillators, the order parameter does not follow the second-order phase transition. Rather we observe an abnormal phase transition of the order parameter due to sudden unbounded trajectories at a critical point. We call it a dangerous aging transition. We provide details bifurcation analysis of such abnormal phase transition. We show that the boundary crisis of a limit-cycle oscillator is at the helm of such a dangerous aging transition.
\end{abstract}
\pacs{Valid PACS appear here}

\maketitle

\section{Introduction}
Studying diffusively coupled oscillators have provided  great insights to elucidate a plethora of  important self-organising activities in various complex systems ranging from physics, chemistry, biology, and  engineering\cite{Kurths-book,Strogatz,kuramoto1984}.  Collective dynamics of coupled oscillators crucially depends not only on the intrinsic dynamics of individual oscillators but also on the nature of coupling topology. In many natural and man made systems robust oscillatory dynamics is an essential requirement for their proper functioning. Emergent rhythmic dynamics of such  large-scale systems should be robust against various local degradation or deterioration.

In recent past,  researchers have spent considerable amount of time to explore the robustness of rhythmic activities in a network  of coupled oscillators when a fraction of the dynamical components are deteriorated or functionally degraded but not removed\cite{daido2004,tanaka2012sr}.  If this degradation reaches a certain critical point, the regular functioning of such systems may hamper and face severe disruption. This emergent behaviour is described as aging transition and is an active area of research\cite{Tanaka2014,Tanaka2015,Tanaka2015a,Tanaka2017,daido2004,tanaka2012sr,liu2016,tanaka2010p,daido2007,PhysRevERakshit17,chandrasekar2019CHAOS}.   The aging transitions might cause catastrophic effects in many natural and real-world systems such as metapopulation dynamics in ecology, neuronal dynamics in brain, cardiac oscillations, and power-grid network\cite{Gilarranz2012,Ranta2008,Lisman2008,Frasaca2017}.  Therefore, it is of  immense practical interest to understand the mechanisms of  aging transition in various complex systems which will eventually help us to propose some control mechanisms to avoid such catastrophes. 

 Till now studies have focussed  on dynamical robustness  by considering different coupling topologies of the network or using different coupling functions\cite{tanaka2012sr,Srilena2019_nonlinear_dyna,Ray_2020,bidesh2020CHAOS}.  Some possible remedial measures have also been proposed by some researchers to enhance the dynamical robustness\cite{Srilena1,Srilena2,Bidesh2019}. Most of  these studies  are based on Stuart-Landau limit cycle oscillator which  describes dynamics near a supercritical Hopf bifurcation. Nevertheless, Kundu et al studied the aging transition in ecological as well as neuronal model\cite{PhysRevERakshit17, Srilena2019_nonlinear_dyna}.  In \cite{PhysRevERakshit17} authors have studied the aging transition in a metapopulation where as in \cite{Srilena2019_nonlinear_dyna}  authors have shown the effects of chemical synapse to enhance the robustness in a multiplex network. To the best of our knowledge till now aging transition has not been well studied for relaxation oscillators. In our present study  we explore how aging transition takes place in a network of Van der Pol oscillators.   The Van der Pol equation  which exhibits non-sinusoidal oscillation  is a well known relaxation oscillator and it  has been used widely to model phenomena in physical, engineering and biological sciences.
  Extensions of such oscillators are very useful  to model the electrical activity of the heart and action potentials of neurons\cite{Sudeshna2014SR}.

 In this work, we have investigated the mechanism of aging transition in a network of a diffusively coupled van der Pol oscillators. Mathematically this situation can be modeled as a network of coupled oscillators where oscillatory nodes switch to equilibrium mode progressively \cite{daido2004}. If the number of nodes that perturbed to equilibrium state from oscillatory state reaches a critical level, the normal activities of such systems may hamper and face severe disruption. We have observed that the characteristic of the aging transition in a globally coupled network of Van der Pol oscillators is very different from that of Stuart-Landau oscillators. Our investigation reveals that as the network undergoes aging transition the dynamics of the whole network become unbounded at a critical point. This is a matter of serious concern in practical systems, and we call it a dangerous aging transition. We have provided details bifurcation mechanisms responsible for such a catastrophic aging transition. We have established the fact that this dangerous aging transition takes place as a result of the boundary crisis of the limit cycle attractor.

\section{MODEL}

For our present study consider the prototypical Van der Pol oscillator with nonlinear damping governed by a second-order differential equation. The mathematical form of this relaxation oscillator is given by
\medskip
\begin{equation}\label{eq:1}
\begin{split}
\ddot{x} + {\mu}(x^2-1)\dot{x}+x = 0\
\end{split}
\end{equation}

where $x$ is the dynamical variable, and $\mu$ is the parameter. For $\mu<0$ it has a stable equilibrium and at $\mu=0$  a limit cycle attractor comes into existence through a Hopf bifurcation. One can observe  relaxation  oscillation for $\mu>1$. Equating $\dot{x}=y$ gives

\begin{equation}\label{eq:2}
\begin{split}
\dot{x} & =y\\
\dot{y}  &= - {\mu}(x^2-1)y-x 
\end{split}
\end{equation}
Here the period of oscillations is determined by some form of relaxation time and it is quite different from sinusoidal or harmonic oscillations. 
The Van der Pol equation given above can be used to model various  real life phenomena in physical, engineering, and biological sciences. 

%Extensions of such oscillators have been used extensively to model the electrical activity of the heart and the action potentials of neutrons.

\section{Network of Coupled Oscillators }

Next we consider a network of $N$ diffusively coupled Van der Pol oscillators represented by the following equation:
\begin{equation}\label{eq:3}
\dot{\mathbf{X}_i} = \mathbf{F(X}_i) + {\kappa}\sum_{j=1}^{N}A_{ij}(\mathbf{X}_j-\mathbf{X}_i)
\end{equation}
where $\mathbf{X}_i = (x_i, y_i)^T $ denotes the state vector and $\mathbf{F(X}_i) = (f(x_i,y_i), g(x_i,y_i))^T$ represents the inherent dynamics of the $i$-th node as in equation \eqref{eq:1} for $i=1, 2, ..., N$.  In our present study we consider $\mathbf{F(X}_i) = (y_i,- {\mu}(x_i^2-1)y_i-x_i )^T$. Which implies that the inherent dynamics of each individual node is given by the Van der Pol oscillator. Second term is the diffusive coupling that describes nature of interactions
between dynamical elements $i$ and $j$.     $A_{ij}$ represents the adjacency matrix where $A_{ij}=1$ if  $i$-th and $j$-th nodes are connected, otherwise $A_{ij}=0$. Here we consider that  a fraction $p$ of nodes are inactive. That means out of $N$ nodes,  $pN$ number of  nodes are in inactive mode(equilibrium state), whereas  $(1-p)N$ number of nodes are in active mode(oscillatory mode). Whenever a fraction $p$ of the nodes become inactive due to local deterioration  we observe phase synchronised oscillation. The individual inactive oscillator, which cannot oscillate by itself, able to oscillate in the network due to continuous inputs from neighbouring active oscillators. To measure the dynamical  activity in the network we define an order parameter $R$, where  $R= \frac{1}{N}\sum_{i=1}^{N}(\langle x_{i,max}\rangle_t-\langle x_{i,min}\rangle_t)$.  Here nonzero $R$ signifies persistence of oscillation in the network whereas $R=0$ implies absence of dynamical activity in the network.

%Next we explore the aging transition while estimating the critical inactivation ratio $p_c$ for homogeneous  as well heterogeneous network.  We study how the critical value $p_c$ varies with coupling strength $k$ for different network topologies.

\subsubsection{Global Coupling}

For our present study we consider global (all-to-all) coupling  and investigate the aging effects. Without loss of generality, we  set the group of inactive elements for  ${j=1, 2,...,Np}$ and rest of the nodes as active for ${j=Np+1,...,N}$. For our numerical simulations, we set $N=200$ as total number of nodes in the network.  For active oscillators we set $\mu=1.2$ and for inactive we set $\mu=-0.1$. Our choice of parameter value is based on the assumption that as the active node becomes inactive(due to perturbation) its parameter value will be in close proximity of  the Hopf bifurcation point.   With these choice of parameter values, an isolated node manifests either a stable equilibrium state at the origin (inactive oscillator) or a stable relaxation oscillations (active oscillator). 

%**********************Figure-1*******************************************************
\begin{figure}[ht]
	\begin{center}
		\includegraphics[scale=0.5]{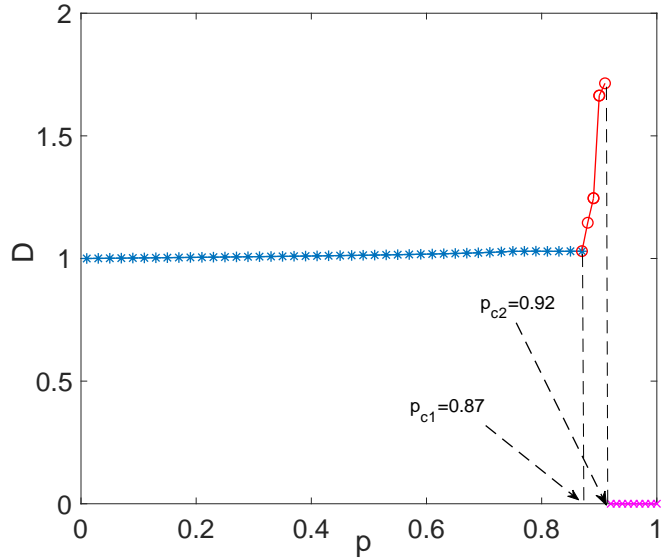}
		\caption{The normalized order parameter $D=\frac{R(p)}{R(0)}$ is plotted against the inactivation ratio $p$ for the coupling strength $k=20.0$.
			\label{fig:fig1}}
	\end{center}
\end{figure}
%*************************************************************************************************
%**********************Figure-2*******************************************************
\begin{figure}[ht]
	\begin{center}
		\includegraphics[scale=0.5]{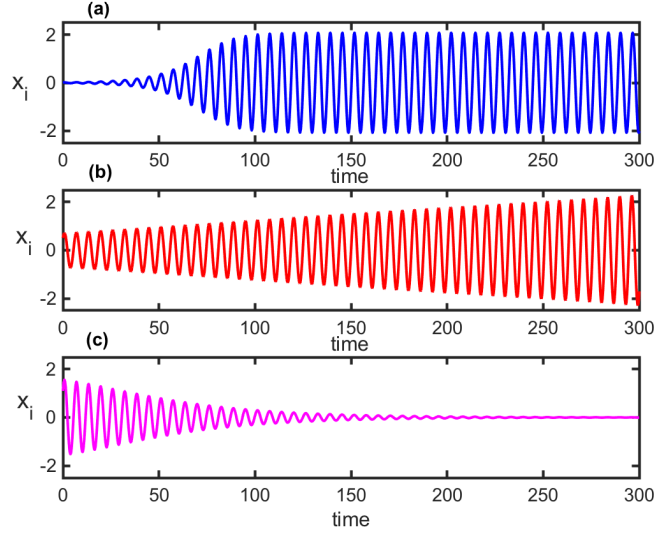}
		\caption{Time series for different inactivation ratios:(a) $p=0.8$, (b) $p=0.9$, and (c) $p=0.95$.
			\label{fig:fig2}}
	\end{center}
\end{figure}
%*************************************************************************************************

For studying the macroscopic oscillation level of the entire network in Fig.\ref{fig:fig1} we have plotted the  the normalised order parameter $D=\frac{R(p)}{R(0)}$ against inactivation ratio $p$ for  the coupling strength $k=20$. There are two critical points  $p_{c_1}$ and $p_{c_2}$ in the figure where explosive phase transitions are taking place.  From $p=0$ to $p=p_{c1}$ the the order parameter remains constant. At the critical point  $p_{c1}$  the order parameter  becomes unbounded. This basically implies that the dynamics of individual nodes diverges to infinity. This scenario continues  till we reach  the second critical point $p=p_{c2}$. At $p_{c2}$ another phase transition takes place  and the order parameter $D$ converges to zero and remains the same till the end point $p=1$.  In the range $(p_{c1}, p_{c2})$ we have considered the finite time length to simulate the network dynamics  in order to show the diverging nature of the order parameter $D$. In support of our claim, in Fig.\ref{fig:fig2} we have plotted the time series of the network for different inactivation ratio $p$. In Fig.\ref{fig:fig2}(a) we have considered $p=0.8<p_{c1}$ and the network dynamics converges to a stable limit cycle behaviour. For $p=0.9$ which is in-between two critical points $p_{c1}$, and $p_{c2}$ we observe diverging time series (Fig.\ref{fig:fig2}(b)) and for $p=0.95>p_{c2}$ network dynamics converges to a stable steady state(Fig.\ref{fig:fig2}(c)). Simulation of the network of oscillators while it is experiencing aging transition reveals two surprising facts. We observe that even though the network is undergoing aging transition the average amplitude of the network remains almost same  till the inactivation ratio $p$ reaches  its critical value $p_{c1}$. Our second observation  is abnormal phase transition at $p=p_{c1}$ which is explosive one and it has a far reaching consequence as far as the practical systems are concerned. This explosive nature of the phase transition might cause great damage to the system.  We call it a dangerous  aging transition. At the same time since the  average amplitude of the system does not give any indication of the impending aging transition and the consequent abnormal phase transition of the order parameter  its is also very difficult to  develop any early warning signal.

%**********************Figure-3*******************************************************
\begin{figure}[ht]
	\begin{center}
		\includegraphics[scale=0.5]{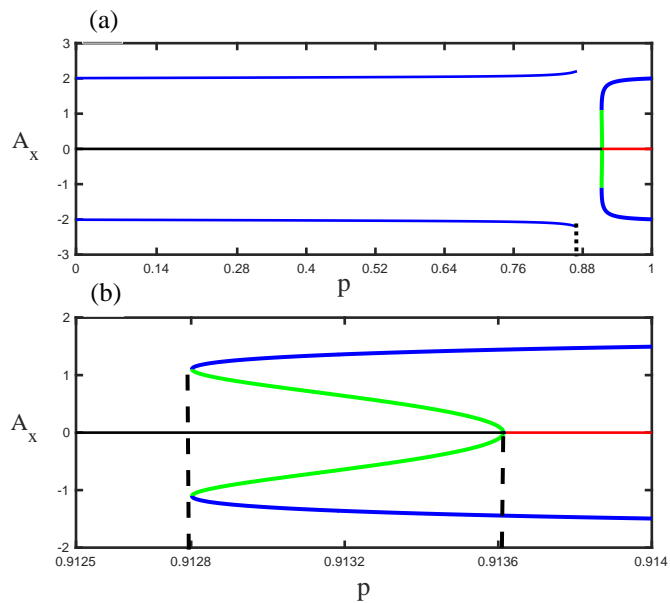}
		\caption{(a) Bifurcation diagram of the reduced system(using XPPAUT) with respect to the inactivation parameter $p$. Here the black line corresponds to an unstable equilibrium while the red line represents a stable equilibrium. Green and blue line represents a stable and unstable limit cycle respectively.
		              (b) Zoomed portion of figure (a).
			\label{fig:fig3}}
	\end{center}
\end{figure}
%*************************************************************************************************

For better understanding of  this fascinatingly surprising phenomena  and some analytical treatment of our results we use the system reduction technique  developed by Daido et al\cite{daido2004}.  Synchronised activity among the oscillators permits us to reformulate the coupled system.  By setting $x_i=A_x$, $y_i=A_y$ for active oscillators and $x_i=I_x$, $y_i=I_y$ for inactive oscillators we reduce the coupled system (\ref{eq:3}) as

%--------------------------
\begin{eqnarray}
\dot{A_x} &=&{A_y}+Kp(I_x-A_x),\nonumber\\
\dot{A_y} &=&-\mu_1(A_x^2-1)A_y-A_x+Kp(I_y-A_y),\nonumber\\
\dot{I_x} &=&{I_y}+Kq(A_x-I_x),\nonumber\\
\dot{I_y} &=&-\mu_2(I_x^2-1)I_y-I_x+Kq(A_y-I_y)\nonumber\\
\label{e5}
\end{eqnarray}
%---------------------------
\noindent  where $q=1-p$. 

To obtain the value of $p_{c2}$ analytically, a linear stability analysis of the reduced system (\ref{e5}) around the origin ($A_x =0, A_y=0, I_x=0,I_y=0$)  can be carried out. The point $p=p_{c2}$ is basically the Hopf bifurcation point of the reduced system. In Fig.\ref{fig:fig3} we have plotted the bifurcation diagram (using XPPAUT \cite{XPP-Book}) with respect to the inactivation parameter $p$ and it  reveals some interesting 
facts about the aging transition. There is a stable limit cycle along with an unstable equilibrium point  till the bifurcation parameter  reaches the the value $p=0.88$.  At this point, the stable limit cycle ceases to exist and the trajectories diverge to infinity.  This dynamical behaviour  of the reduced system perfectly matches with the actual network dynamics. The parameter value $p=0.88$ corresponds to $p_{c1}$ in the actual network. At the $p=0.9128$  a pair of  stable (green) and unstable (blue) limit cycles come into existence through fold limit cycle bifurcation. Finally at the point $p=0.9136$  stable limit cycle disappears through a Hopf bifurcation and  the unstable equilibrium point becomes stable one. Here $p=0.9136$ corresponds to $p_{c2}$ in the actual network. 
%**********************Figure-4*******************************************************
\begin{figure}[ht]
	\begin{center}
		\includegraphics[scale=0.5]{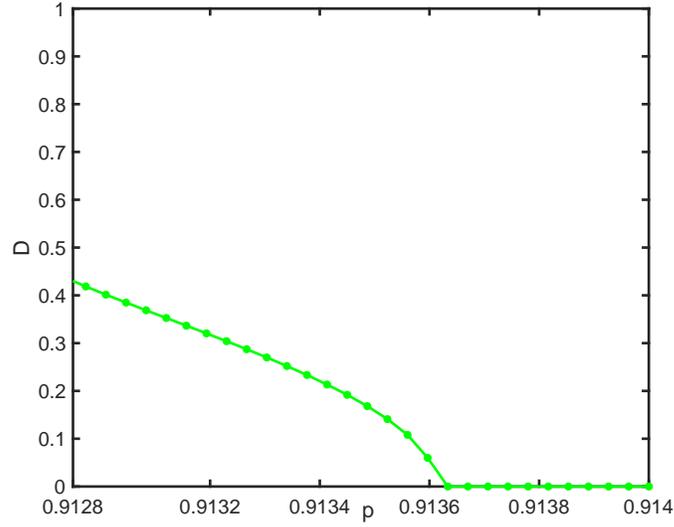}
		\caption{The normalized order parameter $D=\frac{R(p)}{R(0)}$ is plotted against the inactivation ratio $p$  in the range $(0.9128,0.9137)$. Here we consider  $N=10000$. and coupling strength $k=20.0$.
			\label{fig:fig4}}
	\end{center}
\end{figure}
%*************************************************************************************************

The existence of the small limit cycle, which is shown in Fig.\ref{fig:fig3} for the reduced model in the range (0.9128, 0.9136), is not apparent in Fig.\ref{fig:fig1} for the actual network dynamics. This is be because of the small size of the network we have considered due to lack of computational facility. However,  in Fig.\ref{fig:fig4} we have plotted the normalised order parameter $D$ for a small range of parameter  value  $p$  by considering $N=10000$. In this figure one can clearly observe the existence of a small limit cycle in the range $(0.9128,0.9136)$.

\begin{figure}[ht]
	\begin{center}
		\includegraphics[scale=0.5]{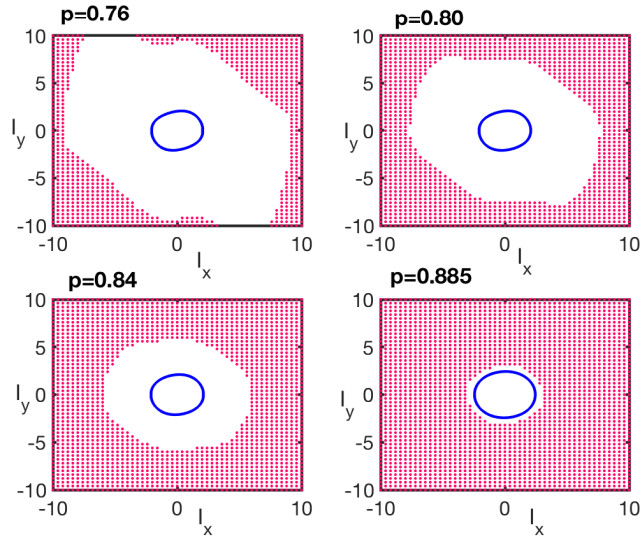}
		\caption{ 2-dimensional cross section of  the 4-dimensional basin of attraction for the reduced system for different values of the inactivation parameter $p$. White region is the basin of attraction of the limit cycle. One can clearly see that the basin of attraction gradually decreases as the parameter $p$ approaches the $p_{c1}$.
			\label{fig:fig5}}
	\end{center}
\end{figure}

 Next we investigate divergence of trajectories after the critical point $p_{c1}$. In Fig.\ref{fig:fig5} by using Poincare section representation we have plotted 2-dimensional cross sections($I_{x}-I{_y}$ plane)  of 4-dimensional basin of attractions of the stable limit-cycle attractor for different values of the inactivation parameter $p$. We have calculated the 4-dimensional basin of attraction by varying $I_{x,y}$ in the range $(-10,10)$ and equating $A_{x,y}=I_{x,y}$. White region represents the basin of attraction of the limit cycle. We have also plotted the limit-cycle attractor(blue circle). From this figure we can observe that as the parameter $p$ tends to the critical value $p_{c1}$ the basin of attraction gradually decreases and the limit-cycle attractor approaches the basin boundary. So it clearly demonstrates that  at the the point $p=p_{c1}$ a boundary  crisis for the limit cycle attractor takes place and this global bifurcation eventually leads to the unbounded  behaviour of the trajectories.

%and the following Jacobian
%matrix is obtained:
%\begin{center}
%J=$\begin{pmatrix}
%-Kp&1&Kp&0\\
%-1&(\mu_1-Kp)&0&Kp\\
%Kq&0&-Kq&1\\
%0&Kq&-1&(\mu_2-Kq)\\
%\end{pmatrix}$.
%\end{center}

%**********************Figure-5*******************************************************
%*************************************************************************************************
\section{Conclusion}
In this paper, we have investigated the aging transition in a network of diffusively coupled Van der Pol oscillators. Our studies claim that the aging transition in coupled Van der Pol oscillators is qualitatively very  different from that of sinusoidal or harmonic oscillations (Stuart-Landau oscillators). In the case of Stuart-Landau oscillators, the order parameter decreases gradually and it follows a typical second-order phase transition. But in the case of Van der Pol oscillators order parameter does remain  a constant over a large parameter range and then undergoes an abnormal phase transition. In particular, it blow-up suddenly at a critical parameter value.  As we vary the parameter further the order parameter becomes finite again. We have it a dangerous aging transition due to the sudden blow-up of trajectories.  By applying the system reduction technique we have explored the bifurcation mechanism responsible for this surprisingly interesting aging transition. Our study shows that the boundary crisis of the limit cycle attractor,  also known as the blue-sky catastrophe, is accountable for this dangerous aging transition.

Our investigation provides significant new insight on the dynamical robustness of complex systems which can be modelled as a network of coupled Van der Pol oscillators. Our study will have a significant impact to invoke broad interests in the community of nonlinear systems as well as in various applications in the fields of science and technology. In the present study, we have considered regular homogeneous (all to all coupling) network. However, many natural systems follow complex network topology and it will be interesting to study the aging transition of coupled Van der Pol oscillators on the top of a complex network.  It will be part of our future  study.

\section{DATA AVAILABILITY}
The data that support the findings of this study are available from the corresponding author upon reasonable request.
\section{Bibliography}
\bibliographystyle{spphys}       % APS-like style for physics
\bibliography{ref_robust}   % name your BibTeX data base

\end{document}